\documentclass{iaus}
\usepackage{graphicx}

\title[PP~~VISIR-VLT images of K\,3-35] 
{VISIR-VLT Images of the Water Maser\\ Emitting Planetary Nebula
K3$-$35}

\author[Blanco et al.]   
{M\'onica W.\ Blanco$^1$, Mart\'\i n A.\ Guerrero$^1$, Luis F.\ 
Miranda$^{2,3}$, \\ Eric Lagadec$^4$ \& Olga Su{\'a}rez$^5$}

\affiliation{
$^1$Instituto de Astrof{\'i}sica de Andaluc\'\i a - CSIC, Granada, Spain \\
  email: {\tt mblanco@iaa.es, mar@iaa.es} \\[\affilskip]
$^2$Departamento de F{\'i}sica Aplicada, Universidad de Vigo, Vigo, Spain \\ email: {\tt lfm@iaa.es} \\[\affilskip]
$^3$ Consejo Superior de Investigaciones Cient{\'i}ficas, Madrid,
Spain \\
$^4$ European Southern Observatory, Garching bei M\"unchen, Germany \\ email: {\tt
  elagadec@eso.org} \\[\affilskip]
$^5$ Laboratoire Fizeau -Universit{\'e} de Nice Sophia
Antipolis - OCA, Nice, France \\ email: {\tt olga.suarez@unice.fr}}
\pubyear{2011}
\volume{283}  
\pagerange{1--2}
\jname{Planetary Nebulae an eye to the future}
\editors{A.\ Manchado, D.\ Sch\"onberner, \& L.\ Stanghellini}
\begin{document}

\maketitle

\begin{abstract}
K\,3-35 is an extremely young bipolar planetary nebula that contains a precessing
bipolar jet and a small (radius 80 AU) water maser equatorial ring. We
have obtained VISIR-VLT images of K\,3$-$35 in the PAH1 ($\lambda$=8.6
$\mu$m), [S\,{\sc iv}] ($\lambda$=10.6 $\mu$m), and 
SiC ($\lambda$=11.85 $\mu$m) filters to analize the mid-IR morphology and the temperature 
structure of its dust emission. The images
show the innermost nebular regions undetected at optical wavelegths
and the precessing bipolar jets. The temperature map shows variations
in the temperature in the equatorial zone and in regions associated
to its jets.
\end{abstract}



K\,3-35 is an extremely young PN with an optical bipolar morphology
consisting of two lobes separated by a dark lane (\cite[Miranda et al. 2000]{Miranda00}). VLA radio observations
of K\,3-35 have revealed a precessing bipolar jet and a small (radius 80 AU) disk traced by water
masers (\cite[Miranda et al. 2001]{Miranda01}; \cite[Uscanga et
al. 2008]{Uscanga08}). 
Since water maser emission is expected to last only 100 yr after the strong 
AGB mass loss ceases
(\cite[Lewis 1989]{Lewis89}; \cite[G{\'o}mez et
al. 1990]{Gomez90}), water maser emitting PNe (H$_2$O-PNe) K\,3-35 can be 
considered among the youngest PNe identified yet.
Observations of H$_2$O-PNe thus provide important insights on the formation
processes of PNe. K\,3-35 is particularly remarkable
because precessing jets and small sized disks are structural components
believed to play a crucial role in the formation of PNe. We have obtained mid-IR images of K\,3-35 to explore the thermal
emission of the dust using color and 
optical depth maps (\cite[Dayal et al. 1998]{Dayal98}), and to compare the 
mid-IR, optical, and radio morphologies to assess its multi-wavelength 
structure. 
%
%

VISIR-VLT (\cite[Lagage et al. 2004]{Lagage04}) images of K\,3-35 in
the N-band PAH1 ($\lambda$=8.6 $\mu$m, $\Delta\lambda$=0.42 $\mu$m), 
[S\,{\sc iv}] ($\lambda$=10.6 $\mu$m, $\Delta\lambda$=0.16 $\mu$m), and 
SiC ($\lambda$=11.85 $\mu$m, $\Delta\lambda$=2.34 $\mu$m) filters were 
obtained (P85, 09/09/2010) with a 0.075$^{\prime\prime}$ pixel scale 
and 19.2$^{\prime\prime}$ FoV. The seeing during the observations was
0.4$^{\prime\prime}$. 
The data were deconvolved using a Richardson-Lucy algorithm.
We also have used \textit{HST} archive images in the [N~{\sc
  ii}] and $R$ filters (Prop.\ ID: 9101, PI: R.\ Sahai) and VLA radio
continuum images at $\lambda$=3.6 cm
(PI: Y.\ G\'omez) adapted from Miranda et al.\ (2001).


\begin{figure}[t]
\begin{center}
 \includegraphics[height=2.7in]{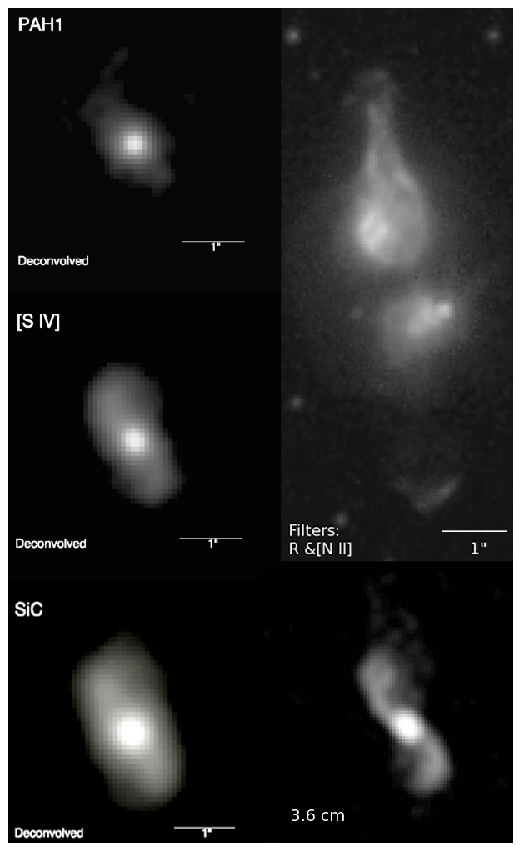}
 \includegraphics[height=2.7in]{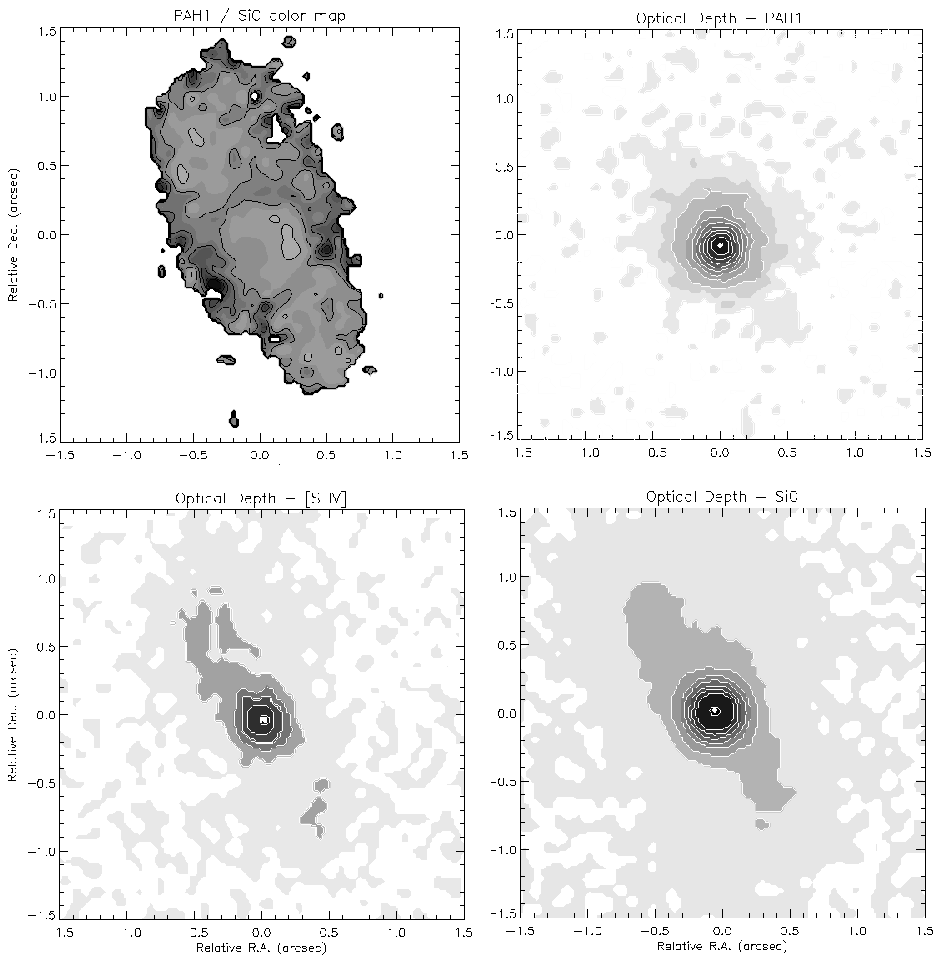}
\caption{
{\it (left)} Deconvolved VISIR-VLT, \emph{HST} [N II] and R, and VLA
3.6 cm radio continuum images of K\,3-35. {\it (right)} K\,3-35 color and optical
depth maps in the N-band.
}
   \label{fig1}  
\end{center}
\end{figure}

The mid-IR emission of K\,3-35 is succesfully detected by VISIR.  
The core, bright in the N-band, is only minimally resolved
with an extension of 0.38$^{\prime\prime}$$\times$0.45$^{\prime\prime}$
along PA$\sim35^{\circ}$. 
The SiC and [S\,{\sc iv}] images show a bright S-shaped structure 
with a size of $2.4^{\prime\prime}$ oriented along PA$\sim30^{\circ}$, which
corresponds to the precessing jets, whereas the PAH1 image show just
the onset of these jets. 
The optical bipolar lobes are only marginally detected in the SiC 
image (Figure\,\ref{fig1}). The comparison of optical, mid-IR and radio images shows that the dark
lane observed at optical wavelengths corresponds to a peak in mid-IR and radio.  
The mid-IR and radio continuum morphologies are very similar to each 
other and clearly show the
precessing jet that are only detected by the bright knots at their
tips in optical images. 
From the 3.6 cm, SiC and [N\,{\sc ii}] images, we infer that dust
emission dominates in the dark lane observed in the optical, while the jets emit mainly in radio continuum, but also in mid-IR. 
The comparison between the SiC and [N\,{\sc ii}] images sets a 
limit for the spatial extent of the dust emission up to the knots at the tips of 
the jets (Figure\,\ref{fig1}).

The PAH1/SIC color map (Figure\,\ref{fig1}) shows variations in the 
dust temperature in the range 230--1550 K
with a mean temperature of 500 K. 
Warm dust at $\sim$500 K traces a ring-like structure around the core. 
The warmest dust is located in the outermost regions of the core along
PA$\sim120^\circ$. Meanwhile, the innermost regions and the S-shaped
precessing jets are cold, although warm dust may be present at the
tips of the jets. 

The optical depth maps (Figure\,\ref{fig1}) reveal larger amounts of material 
at longer mid-IR wavelenghts.  
The emission peaks toward the center of K\,3-35 in all filters, 
implying that the mid-IR emission is optically thin.  
It can be concluded that most mid-IR emission arises from a compact core 
enclosing the H$_2$O-maser-emitting magnetized torus (\cite[Miranda et
al. 2001]{Miranda01}; \cite[Uscanga et al. 2008]{Uscanga08}).

\end{document}